\documentclass{iaus}
\usepackage[dvips]{graphicx}

  \checkfont{eurm10}
  \iffontfound
    \IfFileExists{upmath.sty}
      {\typeout{^^JFound AMS Euler Roman fonts on the system,
                   using the 'upmath' package.^^J}%
       \usepackage{upmath}}
      {\typeout{^^JFound AMS Euler Roman fonts on the system, but you
                   dont seem to have the}%
       \typeout{'upmath' package installed. iaus.cls can take advantage
                 of these fonts,^^Jif you use 'upmath' package.^^J}%
      }
  \else
  \fi

  \checkfont{msam10}
  \iffontfound
    \IfFileExists{amssymb.sty}
      {\typeout{^^JFound AMS Symbol fonts on the system, using the
                'amssymb' package.^^J}%
       \usepackage{amssymb}%
       \let\le=\leqslant  
         
      }{}
  \fi

  \IfFileExists{amsbsy.sty}
    {\typeout{^^JFound the 'amsbsy' package on the system, using it.^^J}%
     \usepackage{amsbsy}}
    {\providecommand\boldsymbol[1]{\mbox{\boldmath $##1$}}}

\newsavebox{\astrutbox}
\sbox{\astrutbox}{\rule[-5pt]{0pt}{20pt}}

\title[Outskirts of Galaxy Clusters: intense life in the suburbs]
{Cluster Mergers, Radio Halos \& Hard X-ray Tails: A Statistical Magneto-Turbulent Model}

\author[R. Cassano \& G. Brunetti]%
{Rossella Cassano $^{1,2}$ \& Gianfranco Brunetti $^2$}

\affiliation{$^1$Dip.di Astronomia, Universit\` a di Bologna
\\[\affilskip]
$^2$Istituto di Radioastronomia del
CNR di Bologna}

\pubyear{2004}
\volume{195}
\pagerange{1--8}
\date{?? and in revised form ??}
\jname{Outskirts of Galaxy Clusters: intense life in the suburbs}
\editors{A. Diaferio, ed.}
\begin{document}

\maketitle

\begin{abstract}
There is now firm evidence that the ICM consists of a 
mixture of hot plasma, magnetic fields and relativistic 
particles. The most important evidences for non-thermal 
phenomena in galaxy clusters comes from the diffuse Mpc-scale
 synchrotron radio emission (radio halos) observed in
a growing number of massive clusters (\cite{Fer03}) and 
from hard X-ray (HXR) excess emission (detected in a few cases)
which can be explained in 
terms of IC scattering of relativistic
electrons off the cosmic microwave background photons
(Fusco-Femiano et al. 2003).
There are now growing evidences that 
giant radio halos may be naturally accounted for by 
synchrotron emission from relativistic electrons
reaccelerated by some kind of turbulence generated
in the cluster volume during merger events (\cite{Bru03}).
With the aim to investigate the connection between
thermal and non-thermal properties of the ICM, we have
developed a statistical magneto-turbulent model which 
describes the evolution of the thermal and non-thermal emission 
from clusters. We calculate the energy and spectrum of the
magnetosonic waves generated during cluster mergers, 
the acceleration and evolution of relativistic electrons and thus the 
resulting synchrotron and inverse Compton spectra. 
Here we give a brief description of the main results,
while a more detailed discussion will be presented in a forthcoming
paper (Cassano \& Brunetti, in preparation). Einstein-De Sitter
cosmology, $H_o=50$ km $s^{-1}$$Mpc^{-1}$, $q_o=0.5$, is assumed.
\end{abstract}

\firstsection 

\section{Introduction}

\noindent Giovannini, Tordi and Feretti (1999) found that $\sim$5\% 
of clusters from a complete X-ray flux limited sample 
have a radio halo source. 
The detection rate of radio halos shows an abrupt increase
with increasing the X-ray luminosity and mass of the host clusters:
about 30-35\% of the galaxy clusters with 
X-ray luminosity larger than 10$^{45}$ erg s$^{-1}$
show diffuse non-thermal radio emission (\cite{Fer03}).
Recent papers (\cite{Ensslin03}; \cite{Kuo03}) 
have investigated the statistics of the
formation of radio halos from a more theoretical point of view. 
In these works, however, the expected statistics are not derived
from formation models of radio halos, but simply from the observed
luminosity-mass correlations and mass thresholds.\\
We model the formation of radio halos and HXR tails
in a self--consistent approach which follows, 
at the same time, the evolution of the thermal properties of the ICM
and the triggering and evolution
of the non--thermal phenomena assuming magneto-turbulent 
re--acceleration of relativistic particles.
In particular, we follow the formation and evolution of clusters of galaxies,
the generation of merger-driven turbulence and magnetosonic waves 
in the cluster volume, the acceleration and time--evolution 
of the relativistic particles, and of the related non--thermal
emission.

\section{The statistical magneto-turbolent model}

\noindent The major steps of the modelling can be sketched as follows :
\begin{itemize}

\item[{\it i)}] {\it Cluster formation}: 
The evolution and formation of galaxy clusters is computed
making use of a relatively simple
semi--analytic procedure based on the hierarchical
Press \& Schechter (1974) theory of cluster formation.
Given a present day mass and temperature of the parent clusters, 
the cosmological evolution (back in time) of the cluster
properties (merger trees) are obtained making use of Montecarlo
simulations. 
A suitable large number of trees
allows us to describe the statistical cosmological
evolution of galaxy clusters.

\item[{\it ii)}] {\it Turbulence in Galaxy Clusters}:
The turbulence in the ICM is supposed to be
injected during cluster mergers. 
The energetics of the turbulence is estimated from 
standard recipes based on {\it Ram Pressure Stripping} 
theory and it is $E_{tur}\sim \rho v_i^2 \pi R_s^2 R_v$ where $\rho$ is 
the density of the ICM, $v_i$ is the velocity between
the two colliding clusters, $R_v$ is the virial radius of the most massive
cluster and $R_s$ is the stripping radius (\cite{Fuj03}).
We assume that a fraction {\it $\eta$} ($E_{MS}=\eta E_{tur}$)
of the energy of the turbulence is associated to 
{\it fast magneto--acoustic waves} (MS waves).
We use these waves since their damping
rate and time evolution basically depend 
on the properties of the thermal plasma which are provided 
by our merger trees for each simulated cluster.
The spectrum of the MS waves (following the injection 
of turbulence) is then calculated at each time step solving a
turbulent-diffusion equation in the wave-number space (\cite{Eil79})
and assuming that the turbulence injected in the cluster volume 
for each merger event is injected for and thus dissipated 
in a dynamical crossing time.
\begin{figure}
\begin{center}
 \includegraphics[width=9.1cm,height=7.4cm]{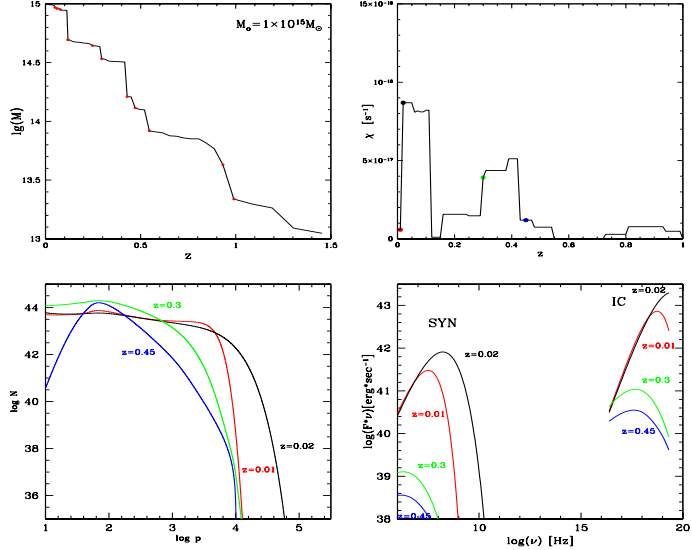}
  \caption{a) {\it Merger-tree} for a cluster with a present-day mass 
$M_{o}=1\times10^{15}M_{\odot}$ ;
b) evolution of the {\it electron acceleration coefficient}
($\chi_{acc}$=$\tau_{acc}^{-1}$, $\tau_{acc}$ is the acceleration time) ; 
c) {\it electron spectra} at four different z;
d) correspondent {\it radio} (synchrotron) and {\it hard-X ray} 
(IC) emission.}   
\label{4p}
\end{center}
\end{figure}
\item[{\it iii)}] {\it Particle Acceleration}:
We assume the presence of relativistic electrons
in the ICM which are injected by AGNs, Galactic Winds, and/or 
merger shocks.
Given the calculated spectrum of MS waves ({\it ii}) and 
the physical conditions in the ICM ({\it i}), we
compute the time evolution of relativistic electrons at each time step
by solving a Fokker-Planck equation 
including synchrotron, IC and Coulomb losses 
and the effect of the electron acceleration due to
the coupling between MS waves and particles. 
\end{itemize}
\noindent In Fig.\ref{4p} we report an example of the results 
in the case of a massive cluster. 
The calculations are obtained assuming 
that the energy injected in relativistic 
electrons is of the order of $\sim 1\%$ of the energy 
of the thermal plasma and a volume averaged $<B>\sim 0.5\mu$G.   
The first important result is that, given these reasonable 
assumptions, the radio 
$(L_{R})$ and hard-X ray $(L_{HX})$
luminosities are in agreement with the observed values
(typically, $L_{R}\in[10^{40}-10^{41}]$erg 
$s^{-1}$ and $L_{HX}\in[10^{42}-10^{44}]$erg $s^{-1}$).

\section{Statistics: Rate of Radio Halos  with Mass}

\begin{figure}
\begin{center}
 \includegraphics[width=5.7cm,height=4.5cm]{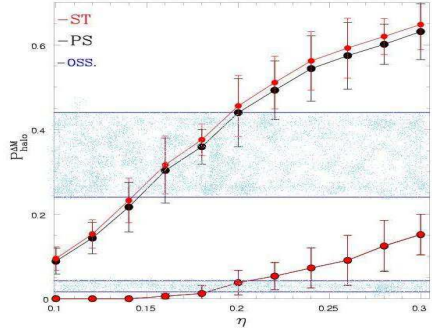}
  \caption{Formation rate of radio halos (points) as a function of the 
parameter {\it $\eta$} in the two mass-bins compared with
observatios (shaded areas): binA (top) and binB (bottom).}  
\label{pp}
\end{center}
\end{figure}
\noindent Given a population of galaxy clusters with  
present day mass and temperature, combining {\it i)-iii)}, 
we follow the cosmological evolution of the non-thermal emission and the
properties of the thermal ICM. We have calculated the formation rate
of radio halos with cluster mass (at z$\le$0.2)
in two bins consistent with those considerated in the observational studies: 
binA$=[1.8-3.6]\times 10^{15}M_{\odot}$ 
and binB$=[0.9-1.8]\times 10^{15}M_{\odot}$. 
In Fig.(2) we report the formation rate of radio halos in the 
two mass-bins as a function of the parameter {\it $\eta$} 
($\eta=E_{MS}/E_{tur}$, Sect.2). The shaded regions mark 
the observed range of formation rate of radio halos 
(Giovannini et al. 1999) in the binA
(top shaded region) and binB (bottom shaded region).  
We find that our predictions match the observations in both 
the mass ranges for reasonable values of {\it $\eta$} $\in[0.16-0.22]$.
In particular, in agreement with observations, we find that 20-30\% of 
clusters in the binA can form a radio halo and that only 
2-3\% of galaxy clusters in the binB have a radio halo.

\end{document}